\documentclass[aps,pre,twocolumn,groupedaddress]{revtex4-1}
\usepackage{amsmath}
\usepackage{graphicx}
\usepackage{hyperref}
\usepackage{breakurl}

\begin{document}

\title{Reverse-engineering biological networks from large data sets}

\author{Joseph L. Natale}

\affiliation{Department of Physics, Emory University, Atlanta, GA 30322, USA}

\author{David Hofmann}
\affiliation{Department of Physics and Initiative in Theory and Modeling of Living Systems,\\ Emory University, Atlanta, GA 30322, USA}

\author{Dami{\'a}n G.\ Hern{\'a}ndez}
\affiliation{Department of Physics and Initiative in Theory and Modeling of Living Systems,\\ Emory University, Atlanta, GA 30322, USA}

\author{Ilya Nemenman}
\affiliation{Department of Physics, Department of Biology, and Initiative in Theory and Modeling of Living Systems, Emory University, Atlanta, GA 30322, USA}

\date{May 24, 2017}

\begin{abstract}
Much of contemporary systems biology owes its success to the abstraction of a {\em network}, the idea that diverse kinds of molecular, cellular, and organismal species and interactions can be modeled as relational nodes and edges in a graph of dependencies. Since the advent of high-throughput data acquisition technologies in fields such as genomics, metabolomics, and neuroscience, the automated inference and reconstruction of such interaction networks directly from large sets of activation data, commonly known as reverse-engineering, has become a routine procedure. Whereas early attempts at network reverse-engineering focused predominantly on producing maps of system architectures with minimal predictive modeling, reconstructions now play instrumental roles in answering questions about the statistics and dynamics of the underlying systems they represent. Many of these predictions have clinical relevance, suggesting novel paradigms for drug discovery and disease treatment. While other reviews focus predominantly on the details and effectiveness of individual network inference algorithms, here we examine the emerging field as a whole. We first summarize several key application areas in which inferred networks have made successful predictions. We then outline the two major classes of reverse-engineering methodologies, emphasizing that the type of prediction that one aims to make dictates the algorithms one should employ. We conclude by discussing whether recent breakthroughs justify the computational costs of large-scale reverse-engineering sufficiently to admit it as a mainstay in the quantitative analysis of living systems.
\end{abstract}


\maketitle

\section{Lay of the land}
\label{lay_of_land}

Biological systems on all levels of organization, from cells to brains and to populations, are comprised of ensembles of interactions among smaller constitutive components~\cite{hartwell1999,proulx2005,zhu2007}. These interactions are typically very specific, and highly coordinated spatially and temporally~\cite{kholodenko2006,deLeon2007,kholodenko2010,dubuis2013,wangX2012}. Involving not just pairs, but also larger groups of components acting in concert~\cite{buchler2003,wangK2009,margolin2010multivariate,ganmor2011,choDY2012,merchan2016}, they are responsible for the rich diversity of complex phenomena and behaviors that make living systems work. Although often prohibitively numerous to model individually (though see~\cite{hlavacek2009}), these components and their corresponding interactions can be represented formally as graphs~\cite{estrada2011}, known colloquially as {\em biological networks}~\cite{kitano2001,alon2003,alm2003,barabasi2004,alon2006,palsson2006,zhu2007,subramanian2015}.

\begin{figure*}[th!]
\includegraphics[width=1\textwidth,angle=0]{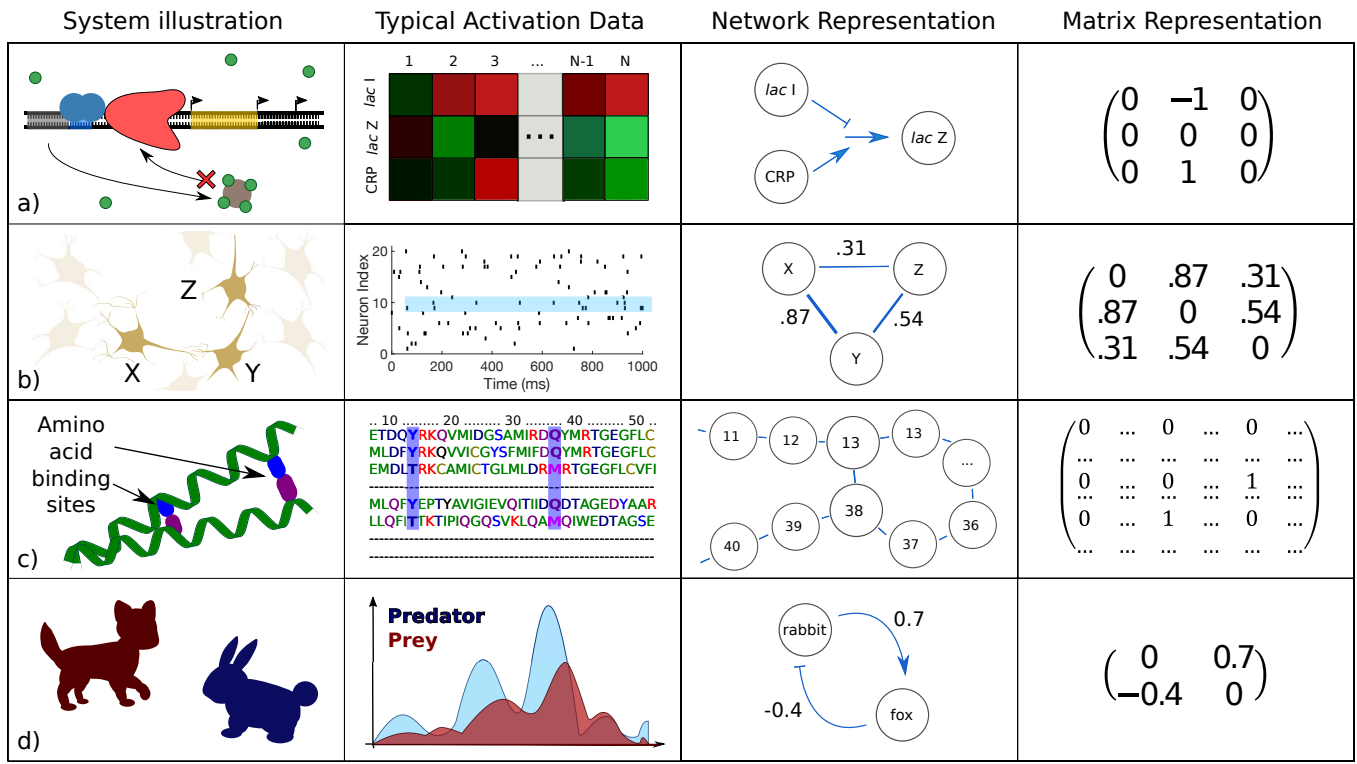}
\caption{\small \label{fig:figure_1}{Examples of biological systems whose constituent interactions have been modeled using networks. (a) The regulation of gene transcription by transcription factors and other enzymes. For example, in the classic \emph{lac} repressor circuit~\cite{jacob1961}, when lactose concentration is high and glucose concentration is low, genes for metabolizing lactose are strongly activated. Activities in this case might consist of microarray data for each mRNA species; the network shows the logic (AND) of the system. (b) Neuronal co-activation networks can be measured by computing correlations between spike patterns. In this case, the network graphs are weighted, and weights may represent correlations. (c) Spatially proximal amino acids tend to co-evolve, as they often participate in bonds that are vital to the structure and function of the protein they form. Here activity values are discrete, assuming one of 20 values to identify the amino acid at each site; the network represents bonds that are inferred to exist by noting which site pairs are highly correlated across similar proteins in different organisms. (d) Complex predator-prey interaction dynamics can be cast in a network form as well. Here activation data represent the populations of each species, and connections are labeled with inferred values for the parameters in the governing population dynamics equations.}
}
\end{figure*}

The variables in such networks (also called nodes) typically represent biochemical or ecological species, cells, or even amino acid residues when one is interested in the biophysics of proteins. The links among the nodes represent interactions, such as chemical transformations, catalysis, and binding; cooperative or predator-prey relations among species; electrical and chemical communication among cells; or geometric proximity among amino acid residues (Fig.~\ref{fig:figure_1}).

To answer many questions in modern data-rich biology, an intermediate step often involves the reconstruction of such networks from empirical data. The data typically consist of joint samples of activities (often referred to as expressions, frequencies, abundances, or population sizes, depending on the context) of a large number of components measured in different biological contexts. Problems of this kind pervade the quantitative life sciences on all physical scales, even if they take different forms and use different languages across scientific disciplines.

At the smallest scale is the problem of inference of physical contacts for amino acids in a protein fold~\cite{morcos2014,halabi2009,feinauer2016}, which is a network representation of the 3D structure of the protein. Predicting such networks from the co-occurence of amino acids promises the ability to design proteins with specific functional properties. At the cellular level, different genes activate or suppress the activities of other genes, forming networks of genetic regulatory interactions~\cite{bansal2007,linde2015}. Similarly, metabolites transform into each other, catalyzed by various enzymes; these form metabolic networks~\cite{palsson2006,cakir2009,cakir2014}, as well as networks that combine both protein and metabolic modalities. Protein signaling networks characterize the structure of decision-making and information processing in individual cells~\cite{kholodenko2002,stark2003,prill2011,Cheong2011,ni2011}. The accurate reconstruction of different types of these cellular networks is expected to lead to successful interventions that cure some of the most debilitating diseases~\cite{little2014}.

On the scale of the nervous system, one often reverse-engineers neural circuits~\cite{schneidman2006,bettencourt2007,friedrich2013,chung2013,epp2015} and, on a larger scale, functional connectivity networks  between brain regions~\cite{stevenson2008,bullmore2009,bullmore2011,friston2011,capone2015}. The structure of the latter has been shown to be valuable as a diagnostic tool for some psychiatric diseases~\cite{lynall2010}, and there is mounting evidence that the former can be ``reprogrammed'' via external interventions to repair damaged circuits~\cite{canter2016}. Finally, on the largest scales, one can reconstruct networks of interactions among members of a particular species~\cite{eagle2009,angluin2010,psorakis2012}, or different species in an ecosystem~\cite{may1971,may1972,may2006,mcCoy2011,sugihara2012,deyle2016}. This knowledge may help in forecasting ecological catastrophes~\cite{moon2015,compton2015} and addressing the spread of infectious disease~\cite{nikolopoulos2016} (or other epidemics~\cite{bahr2009}).

In all of these fields, data share similar properties, and data sets often have similar sizes. This imposes uniformity not only on the question of network inference itself, but also on the obstacles and algorithmic approaches that underlie reconstruction efforts across multiple biological domains. Inference methods designed for one system type~(\cite{arkin1995},~\cite{hofmeyr1996}, and~\cite{margolin2006}) can often be adapted to accommodate others (\cite{schmitt2004,dunlop2008,villaverde2014MIDER},\cite{delaFuente2002,kholodenko2002,santos2007}, and~\cite{bettencourt2007,bandaru2011}, respectively). Moreover, morally equivalent methods have been developed in nominally unrelated fields~\cite{stein2015} -- or else borrowed explicitly from established disciplines, such as systems identification techniques migrating to network biology from engineering~\cite{yeung2002,gardner2003}).

An additional reason for the cross-pollination among the subfields of biological networks inference is that, like in other parts of bioinformatics, the field has benefited from advances in machine learning and related Big Data computational tools. In their turn, as is often true of mathematical approaches, these tools are applicable across multiple traditional biological subdisciplines, and hence provide for natural theoretical bridges not only among life-sciences subfields, but also to a ``network'' of other quantitative disciplines (physics, statistics, and computer science)~\cite{villaverde2014Royal}.

However, one cannot embrace the unembraceable. Thus in this review, we will focus almost exclusively on applications of networks inference to the systems biology of the cell~\cite{christensen2007,vidal2011,Quo2012}, and will mention bridges to other fields only briefly and haphazardly, leaving the reader certainly thirsty for more. Starting with a few of the references that we mention, as well as using Google Scholar (another network, this time of citations), is an easy way to quench this thirst! 

Before proceeding any further, it is certainly worth warning the reader that the explosive growth of the field of biological network inference has covered with a thick blanket of journal articles some treacherous rocks. A few of them are very dangerous, and can, in principle, sink the field if not addressed thoughtfully. Specifically, while fully automated network inference has become a routine procedure, it is not immediately clear that the large-scale reconstruction of entire networks from high-throughput data will necessarily result in tangible insights or actionable understanding about biological systems. One reason is that most reconstructions are not experimentally verified, remaining in the literature as collections of information (or misinformation) of dubious quality. Another comes from the fact that it is still not clear what new knowledge entire-network inference yields, besides proposing potential interactions for experimental verification. If a goal of the field is to predict response of biological systems to yet-unseen exogenous perturbations, then the bridge between a network graph and such predictive knowledge will have to be built eventually, but it is not there yet in most practical applications. Most importantly, it is usually unclear what insights are delivered by large-scale networks, or how to interpret the typical product of the reconstruction enterprise -- Lander's infamous ``hairball'' of decontextualized interactions~\cite{lander2010}. One can even argue that exhaustively enumerating interactions is not inherently more insightful than cataloging the original experimental data, and both should give way to studying the system's emergent properties~\cite{anderson1972}. Having now warned the reader, we leave these important, foundational questions aside for the remainder of the review (save the \emph{Discussion}).

\subsection{Scale of the biological network inference problem}
\label{scale_of_problem}

{\em Network reverse-engineering is typically done in the ``low-hanging fruit,'' Big Data regime. Here the data sets are large, but the number of unknowns is even larger: not all the unknowns can be learned reliably.}

\vspace{0.1in}

While reconstructions can be performed using different data types \cite{hecker2009,christensen2007}, here we are concerned with approaches that are based exclusively on biological {\em activity measurements}. Suppose we have a network consisting  of $p$ nodes (e.~g., a group of $p$ interacting genes or neurons), and $n$ simultaneous measurements of some activity variable for each of these nodes (which for our purposes fully characterizes the biological states of the nodes at a given moment in time). The activity variables can be binary (as in the characterization of whether a gene is on or off, or whether a neuron is spiking or not at a given time) or real-valued (gene expression levels, or firing rates for neurons). In other words, the total amount of available data is $\sim np$. The goal is to identify links between pairs of the $p$ nodes (or more generally, higher order interaction structures) from patterns in their activities. If we focus on pairwise interactions among the nodes only, then the number of unknowns is  $\sim p^2$. Thus the amount of data per unknown is $\alpha\sim np/p^2=n/p$.

In the classical statistics regime, the amount of data is typically asymptotically large compared to the number of unknowns, $\alpha\gg1$. In contrast, network inference usually proceeds in the regime where  $p\gg1$, with typical $p\sim 10^2\dots10^3$. For gene expression and other high throughput cellular data, in particular, it is not uncommon to have $p\sim 10^4$. Other fields are catching up~\cite{ahrens2013,schwarz2014}.  The number of measurements is also typically large, $n\gg1$. We can consider $n<p$, as in most genetic data, or $n>p$ (but not $n\gg p$), as in many neuroscience applications. More generally, $n\sim p$, so that $\alpha\sim1$, representing a qualitative departure from the classical statistics regime.

The situation gets even worse when we remember that the total number of parameters characterizing all (higher-order) interactions in a network scales as the total number of states that the network can be in (i.~e., $2^p$ for binary nodes, or $2^{pS}$ for continuous ones, where $S$ is the entropy of each node measured at the experimental resolution). Thus in the most general case, for biological network inference, $\alpha\ll1$. It is clear then that, just like in most other Big Data applications, the problem cannot be solved completely, with all interactions identified. Thus networks inference necessarily is a ``low-hanging fruit'' problem, where the limited data allows us to focus only on the most salient features of the studied systems. This also means that, in any quantitative assessment of the quality of network reconstruction methods, we should focus a lot more on the precision (absence of false positives) of a method, rather than on its sensitivity (absence of false negatives), since the sensitivity of essentially any method on realistic data would be tiny.

\subsection{Different ideologies for inference}
\label{mech_vs_eff}

{\em In biological network inference, one can think of reconstructing actual physical interactions among the nodes or coarse-grained, phenomenological surrogates. We focus exclusively on the latter.}

\vspace{0.1in}

The notion of network inference may evoke the idea of reconstructing actual physical interactions among network nodes. For example, a regulatory interaction between two genes might mean the direct binding of a transcription factor protein, translated from one of these genes, to a specific part of the DNA sequence that controls the expression of the other gene~\cite{dunlop2008}. We refer to the reconstruction of such physical, microscopically accurate interactions as the inference of \emph{mechanistic} networks. In contrast, the majority of reconstruction methods focus (explicitly or not) on the inference of \emph{effective} interaction networks~\cite{timme2014}, which keep track of purely phenomenological interactions. These may or may not be mechanistically accurate, but are sufficient to reproduce various statistics of the observed variables. Such effective interactions may correspond to subsets of the interactions in mechanistic networks. They may be compact, coarse-grained averages of some microscopic quantities. Or they may be entirely macroscopic properties that have remote and complicated relationships with the microscopic, mechanistic interactions.

One can focus on effective network inference for purely pragmatic reasons: as discussed above, even high-throughput data is insufficient to infer \emph{all} the contributing actors in a complex system, and effective interactions may simply be the low-hanging, accessible fruit. In contrast to this pessimistic view, one may argue that every level of description requires its own proper degrees of freedom for efficient representation~\cite{anderson1972,goldenfeld1999,daniels2015}, and that the distinction between mechanistic and effective networks is not that clear-cut.

To wit, even mechanistic biophysical interactions are themselves effective interactions, just at a different scale. For example, the bonds between amino acids that form at protein-protein interfaces consist of electrostatic forces between constituent molecules. These forces can then be broken down in terms of quantum interactions between elementary particles, at which point the notion of an amino acid has long since disappeared. Likewise, the fact that communication between synapsing neurons requires the diffusion of neurotransmitters across the synapse undermines the notion that neurons can ever truly be in a direct, mechanistic contact. We are most sympathetic with this viewpoint, which treats the distinction between mechanistic and effective networks less as a dichotomy than as a spectrum. In what follows, we cast the issue in terms of modeling assumptions: what is the appropriate set of nodes and interactions to answer {\em the specific questions being asked} while working at the {\em desired scale}?

Our perspective notwithstanding, a few authors have distinguished explicitly between these two ideologies (see~\cite{gardner2005} as the originator of the ``physical'' vs. ``influence'' network terminology, and~\cite{friston2011} for a more fine-grained distinction among different types of effective networks in the brain). Many other sources refrain from making such explicit distinctions, presumably either for expedience in exposition or because they take seriously the aforementioned notion of pursuing the most efficient or useful description at a given level of study, regardless of the biological implementation details at other levels. While we remain agnostic to the particular reasons for the tendency of reverse-engineering literature to avoid making this distinction at the outset, we lament the absence of explicit declarations of the intended level of description when elaborating a new algorithm by the majority of publications. By default, in this Chapter, we focus on effective inference methods, for which authors do \emph{not} make an effort to understand whether there is a mechanistic basis for inferred interactions, stating any exceptions at the outset when they appear.

\subsection{Goals of this Chapter}
\label{goals}

We are now in a position to state our intended goals for this Chapter. In the following sections, we review relatively recent (within the last two decades) attempts at network inference, contending:

\begin{enumerate}
\item The aptness and success of a given inference method depend on the ultimate purpose of performing network reconstruction. One must first establish what kinds of predictions are desired (i.~e., what does one seek to \emph{learn}~\cite{alm2003} using the network?), and only then decide which algorithm to use.
\item Large-scale network reverse-engineering has many fruitful applications, but it is not always the necessary -- or not necessarily the best -- approach for making certain kinds of predictions.
\end{enumerate}

Note that we deal exclusively with inference methods that produce networks containing at most pairwise interactions. While the joint probability distribution for $p$ discrete or continuous stochastic activation variables in a stationary state $\lbrace g_i \rbrace$ can be expanded~\cite{margolin2006} most generally as
\begin{multline}
P(\lbrace g_i \rbrace ) \propto \exp \left[ -\sum_i^p h_i (g_i) -\sum_{i,j}^p J_{ij}(g_i,g_j)\dots\right. \\ \left. -\sum_{i,j,k}^p \phi_{ijk} (g_i,g_j,g_k) - ... \right],
\label{expansion}
\end{multline}
where functions $h_i$, $J_{ij}$, and $\phi_{ijk}$ denote first-, second-, and third-order interactions, respectively, it is clear from the considerations of Section~\ref{scale_of_problem} that reliable estimation for terms of higher order than $J_{ij}$ is prohibitively difficult. In addition, we review only the algorithms that attempt to infer \emph{static} values for $J_{ij}$ under the assumption that the system is in (near-)stationary conditions, although some authors have attempted to estimate networks whose topologies are dynamically evolving~\cite{parikh2011,song2009}.

\vspace{0.1in}

The progression of the Chapter is as follows. First, we examine highlights of the many places where network inference has been used to advance new knowledge in contemporary systems biology and establish novel paradigms in modern medicine. Then we proceed to delineate and explicate several types of inference methods, briefly describing the operation of several representative algorithms for each of the classes we name. We conclude with a brief outlook of where the field might be headed. However, these concluding comments should be taken with a lot of caution, since ``it is difficult to make predictions, especially about the future.''

\section{Roles for reverse-engineering in systems biology research}
\label{roles}

\emph{The reverse-engineering of large-scale networks by means of automated algorithms has become such a routine procedure that it
has spawned a research field of its own. Why is the task of learning networks from data considered so important?}

\vspace{0.1in}

The modern imperative to generate comprehensive parts lists for large biological systems~\cite{palsson2006} is epitomized in what one author somewhat flippantly calls ``the giant maps of metabolic pathways that many molecular biologists pin to their walls''~\cite{newman2003}. Such diagrams encode and illustrate visually the entirety of observable interactions of a particular type in a specific system. Since the mid-2000s, attempts to generate such maps have been pursued vigorously by researchers in various disciplines, but the most prominent and systematic efforts have come from the network inference Challenges of the Dialogue on Reverse-Engineering Assessment and Methods (DREAM) initiative~\cite{stolovitzky2007,stolovitzky2009}. Contestants participating in these ongoing Challenges submit network reconstructions, inferred by original algorithms operating on standardized data, for comparison against (experimentally) established sets of interactions in benchmark networks.

The top-scoring networks in early competitions achieved respectable accuracies, despite the difficulties associated with defining ``gold standard'' benchmarks and evaluation metrics~\cite{stolovitzky2007,boutros2014}. However, they also lacked the ability to provide intuition (beyond structural insights) about the systems they described. As static pictures of interaction architectures, they had limited ability to \emph{predict} a system's behavior. The pattern of assembling a large, intricate network as the end goal, with no intention to use it as a tool for prediction -- as in the iconic but largely uninformative hairball of Ref.~\cite{lander2010} -- thematized DREAM competitions roughly until 2014, nearly a decade after one reviewer declared the field to be ``still in [its] `natural history' phase''~\cite{proulx2005}.

The emphasis of DREAM competitions has since shifted, mirroring changes in the attitude of the reverse-engineering community as a whole. Recent competitions have more strongly favored \emph{predictive modeling}, with inferred networks serving not as ends in themselves, but as coarse summaries of high-dimensional data  -- a special type of statistic -- to aid in projecting how the behavior or components of a system will change (as a function of time, due to changes in its environment, etc.).

This movement away from using learned topologies as a signal that the ``work is done,'' and instead toward viewing the entire process of network inference as an intermediate step in an fully-fledged research pipeline~\cite{yamane2016}, is also supported by theoretical work. In particular, it has been argued that structure alone provides insufficient information to achieve an adequate degree of control over the underlying system's dynamics~\cite{gates2016}. In fact, the object of interest is not always a network's structural complexity (density of connections), but its \emph{dynamical} complexity (the number of fixed points it can accommodate), which depends on other parameters beyond structure, such as its connection strengths. Indeed, only the latter is closely tied to the viability of a network architecture in the context of evolution~\cite{tikhonov2016}.

The field's transition -- from descriptive to predictive -- is a natural one, and indeed reminiscent of the progression in other branches of science. While it is not completely clear why there was this prolonged period of exploration without modeling, it is plausible that reverse-engineers first needed to convince themselves that (1) networks can, indeed, be accurately reconstructed from activity data alone, and (2) the achieved reconstructions are statistically significant and reproducible. Furthermore, experimental tools for administering systematic perturbations to the networks under study took a while to develop, so that the need to predict dynamical responses to perturbations had not emerged for a while.  As confidences in the statistical power of reverse-engineering grew, and new experimental tools were developed, the next level of questions naturally emerged. It is in answering this next level of questions that network reconstructions have found their broad spectrum of highly nontrivial, often unique, and even central roles in modern systems biology. For the remainder of this section, we survey several key application areas, focusing on the most impactful types of predictions that reconstructions are capable of generating.

\subsection{Predictions regarding individual nodes or interactions}
\label{individual_predictions}

\textit{Reconstructions can help identify intervention targets or functionally similar cohorts of biological species.}

\vspace{0.1in} 

The advent of modern, high-throughput data acquisition techniques transformed the enterprise of network reconstruction from a painstaking, often collaborative process into an exercise in algorithmic design. Verifying the existence of a single interaction no longer demands corroboration by multiple independent research efforts, and connections can now be inferred in parallel directly from a single set of data. An oft-cited consequence of this change of pace contends that modern reverse-engineering dramatically increases the rate at which hypotheses about potential interactions can be generated. To this end, whole-network reconstructions allow us to rapidly elucidate both the presence and nature of individual interactions, as well as predict the function of individual nodes from knowledge about their neighbors~\cite{yu2013,mulder2014}.

Inference methods designed for the express purpose of proposing novel interactions for experimental verification~\cite{delaFuente2004,margolin2006} have confirmed previously established gene targets~\cite{gardner2003} and identified novel targets for known transcription factors and drugs~\cite{basso2005,csermely2013}. Known broadly as statistical or \emph{association} methods (see ``Who talks to whom,'' Section~\ref{who_talks}), algorithms in this class have also discovered entirely new interactions~\cite{basso2005,faith2007,rao2007,wangK2009,margolin2010multivariate,altay2010,kaleta2010}, with previously unknown gene interactions often being verified experimentally~\cite{kumari2012,ma2014}. In a multi-algorithm litmus test, several of these methods were capable of inferring missing links in artificially corrupted, incomplete versions of established pathways~\cite{meyer2014}.

Network-based strategies for the prediction of protein function~\cite{sharan2007} generalize more traditional approaches, such as clustering analyses~\cite{yu2013}, that have been used to classify genes and proteins according to their role at either the physiological or the network level. Individual gene clusters correspond to distinct functional groups in some systems~\cite{wen1998}. They can be used to infer roles for unclassified elements according to the guilt-by-association (GBA) heuristic (i.~e., assigning functions similar to those of nearby neighbors in the interaction space).

Clustering alone cannot produce a full interaction map, and its applicability is limited because its underlying assumptions are not universal among biological system types~\cite{wolfe2005,gillis2012} (GBA may be more valid for protein-protein interactions than gene-gene interactions, since the latter entail more latent or intervening steps). Nevertheless, clustering is still useful in modern reverse-engineering, predominantly in the data-processing phase that often precedes the inference of full interaction architectures~\cite{bonneau2006}. Clustering the data prior to inference greatly restricts the search space by providing an effective prior to bias the set of candidate interactions. On the other hand, the same idea can also be used to \emph{coarsen} inferred networks: ``module-based'' inference techniques~\cite{deSmet2010} have identified entire groups of genes that are functionally related~\cite{segal2003}. We will return to this idea of identifying coarse functional and conceptual (as opposed to simply structural) units in the {\em Discussion}.

\subsection{Insights from the statistical properties of network ensembles}
\label{statistical_ensembles}

\textit{Certain structural statistics differentiate real biological systems from other kinds of complex networks.} 

\vspace{0.1in}

While the rapid verification of microscopic interactions undeniably constitutes an improvement in the pace of discovery, it does not by itself generate categorically new kinds of knowledge. Systems biology is ``more than an accelerated program of molecular biology''~\cite{lander2010}, and the relatively new tools of reverse-engineering must prove their worth by helping to play a part in that grander enterprise. This is reflected in the possibility of using reconstructions to make predictions not only about single nodes and individual connections, but about the statistical properties of network \emph{ensembles}.

Work in this direction has produced various insights about what distinguishes biological systems and endows them with their unique characteristics among complex networks. For instance, it has been shown that the most highly connected nodes in protein networks are likely to be essential~\cite{kravchenko2012} for survival \cite{jeong2001,he2006}. Moreover, nodes with an exceptionally high degree (i.~e., number of connections), called \emph{hubs}, attach preferentially to nodes with low degree while tending to avoid one another~\cite{maslov2002}. This property, in part, underlies the widely observed \emph{modular} organization of cellular systems: an efficient coding scheme in which network partitions include only components involved in related processes. This discourages overlap and ensures that (on average) no single node participates in too many processes~\cite{Cheong2011}. This forms the basis for one type of biological robustness~\cite{lemke2004}.

Certain modular structures recur with disproportionately high frequencies in biological systems (with respect to their chance rate of appearance in a random graph~\cite{estrada2011}). Known as \emph{motifs}~\cite{alon2006,shen2002network}, they can endow the network with vital control and design features, such as positive or negative feedback, and are often conserved throughout evolution~\cite{milo2002,sharan2005,stuart2003}. Studying the appearance rates of different motifs across different networks can help clarify the functional ``purpose'' they satisfy within a given network.

While a node's degree is its most fundamental attribute, studying other network parameters has also led to key insights. The betweenness centrality~\cite{estrada2011} for nodes in protein interaction networks has been observed to be even more highly correlated with protein essentiality than their degree~\cite{hahn2005}. Moving beyond individual nodes, it has been argued that the full degree \emph{distribution} is approximately scale-free~\cite{barabasi2004} for many systems, providing deep architectural support for the robustness of biological systems to noise and perturbations, at both environmental and genetic levels~\cite{kitano2004} (yet see~\cite{stumpf2012} for a cautionary note about the associated power-law distributions).

In \emph{network medicine}~\cite{barabasi2011}, clinically relevant predictions can often be made from such high-level statistics, irrespective of whether interactions can be enumerated exhaustively or determined at a fine-grained level. For instance, the aforementioned correlation between a node's degree and its essentiality for survival begets the notion that candidate drug targets can often be ruled out immediately if they are too highly connected, such that using them risks compromising the rest of the network~\cite{lecca2013}.

While one should not focus exclusively on the architectural aspects of dynamically engaged networks~\cite{tikhonov2016}, even microscopic statistics can sometimes go beyond structure to tell rich stories about the behavior of the underlying system. Maximum Entropy~\cite{jaynes1957,Jaynes1982} methods~\cite{stein2015} (see Section~\ref{who_talks}) have been used to learn the effective coupling constants that connect neurons in the retina~\cite{schneidman2006,tkacik2013}, where the inferred values suggest that these networks naturally reside in the neighborhood of a \emph{critical point} in their parameter spaces~\cite{beggs2003}. This might afford such networks an essentially optimal capacity for stimulus representation, as well as information storage and transmission~\cite{beggs2008,shew2013} (though see~\cite{schwab2014} for an alternative viewpoint). For the amino acid interaction networks that keep track of where bonds form during protein folding, the same methods corroborate the idea that geometrically proximal residues tend to coevolve~\cite{feinauer2016} by showing that bond locations can be identified using a simple statistic on the ensemble of viable protein sequences (in this case, correlations between the activations of site pairs).

\subsection{Using statistics to characterize or classify individual networks}
\label{individual_stats}

\textit{Ensemble statistics can help identify defective or emergent properties in a network.}

\vspace{0.1in}

Sometimes, statistical surmises can be used to make statements about the typicality of a particular network. An approach known as \emph{differential networking} (so named to contrast with \emph{differential expression}, a popular type of approach to activation data in gene networks) has been increasingly used for this purpose.

For example, Refs.~\cite{delaFuente2010,ideker2012} discuss the idea of using topological characteristics to solve supervised classification problems, such as determining whether a given network comes from a healthy or a pathological organism. This possibility is explored explicitly in~\cite{lynall2010}, which nominates several criteria (reduced clustering and ``small-worldness,'' reduced probability of high-degree hubs, and increased robustness) as those which are markedly altered in patients with schizophrenia. The reconstruction method developed in~\cite{grechkin2016} was able to identify genes that are either known tumor drivers, associated with biological processes relevant to disease, or correlated with patient prognosis for various types of cancer by examining how pathological networks differ from their counterparts in ``normal'' tissue. Changes in hub structure have also been used to forecast the survival outcome for breast cancer patients~\cite{taylor2009}.

It is worth pointing out that the aforementioned Maximum Entropy methods~\cite{stein2015} provide, in some sense, a complementary approach to ensemble statistics. Rather than addressing only aspects that networks have in common (or can be averaged over), these approaches are predicated on exploiting the intrinsic \emph{variability} at the micro-scale in an attempt to reproduce what is conserved at the macro-scale. This is especially useful wherever diverse microscopic network connectivity structures  are known to produce indistinguishable behavior at coarser resolutions, as in protein folding: there is no one-to-one mapping between amino acid sequence and tertiary structure, but an entire distribution of microscopic parameters -- a wide variety of equally viable amino acids sequences -- that code for roughly the same protein shape~\cite{weigt2009,marks2011,colwell2014,morcos2014}. Knowing this, one can easily imagine how running Maximum Entropy methods in reverse can help determine, for example, whether a given amino acid interaction network represents a viable protein. The same might be said for evaluating the typicality of an inferred retinal network, by measuring properties like criticality~\cite{shew2009,tkavcik2015} (NB: for a selection of competing viewpoints on the criticality of neuronal networks, consult the aforementioned~\cite{schwab2014}, as well as~\cite{chialvo2010,mastromatteo2011,mora2011,beggs2012}).

\subsection{Predicting how a given network will respond to perturbations}
\label{perturbations}

\textit{Reconstructions help identify and quantify response patterns in novel conditions.}

\vspace{0.1in}

Network models capture and summarize complex dependencies the among states of biological components, often allowing one to predict how a system will change its state or behavior with changes in the biological environment (i.~e., modifications affecting the state of one or more nodes or interactions). Commonly studied perturbations can be local~\cite{pinna2010} (e.~g., knockout of a single gene, as in the simulation of deleterious mutations), multifactorial (affecting many elements)~\cite{jansen2003}, or fully global~\cite{emmert2013} (applying a drug to slightly suppress the firing of all neurons in a circuit), and the system's responses can be investigated at local or global levels as well. For instance, one might inquire about the effect of a drug or a mutation on the expression of a single gene, or the success or failure of signal propagation from start to end through a perturbed pathway.

The types of responses that are interesting to researchers vary widely, and range across a spectrum of detail. The simplest and the coarsest entail qualitative predictions: for example, is the activation state of a given node affected by a specific perturbation? Progressing to a more quantitative picture, one can try to predict the actual post-perturbation values for affected nodes, as in the prediction of gene expression levels following a knockout event~\cite{ud2016}. At the finest granularity, models incorporating time-series measurements can be used to forecast the transient behavior for such a gene as it approaches a new steady-state expression level.

Recent DREAM Challenges have provided a testing ground for algorithms aiming to make these types of predictions. The DREAM4 Predictive Signaling Network Modeling Challenge~\cite{eduati2010} instructed contestants to predict phosphoprotein measurements ``using an interpretable, predictive network''\footnote{\url{http://dreamchallenges.org/project/dream4-predictive-signaling-network-modeling}. The solution presented in~\cite{eduati2010} infers a network using Boolean truth tables,  one of the most popular approaches during the early stages of automated network inference~\cite{Liang1998}. This approach has since fallen out of favor, yielding to the more sophisticated methods we discuss in Section~\ref{reconstruction_methods}, but Bayesian networks are often still discretized to Boolean values for convenience.}, and the bonus round of that year's \emph{in-silico} Challenge~\cite{greenfield2010,huynh2010,pinna2010} asked competitors explicitly to predict the system's responses to ``novel'' perturbations that were not encountered in the training data. The DREAM7  Network Topology and Parameter Inference Challenge~\cite{meyer2014} specified the prediction of perturbation outcomes using gene regulatory network models as a separate step from inferring their topologies.

As we discuss later, prediction of time-course trajectories requires directed networks, but the converse is not true: directional links can sometimes be inferred from static data. On the level of qualitative predictions, the linear dynamical systems approach of~\cite{gardner2003} was able to deduce the targets of novel perturbations in a system of nine genes using only steady-state values of their expression levels, following a series of highly controlled perturbations (and the knowledge of which genes were targeted during the perturbations). We consider this result to be particularly important, for two reasons. First, it challenged previously expressed (and still later-held~\cite{wagner2004}) ideas by successfully determining a directed network, despite the fact that the applied perturbations elicited statistically significant changes in the activations of all nodes. Second, later improvements extended the abilities of the algorithm therein to determine which species were ``hit'' by applied perturbations even \emph{without} specifying as inputs which genes were targeted during the data acquisition phase~\cite{diBernardo2005}, reinforcing the idea that $M$ static, independent, but carefully selected perturbation measurements can substitute for a series of time-course measurements taken at $M$ intervals~\cite{bansal2006}.

\subsection{Representing the joint probability distribution for observables}
\label{probability_graphs}

\textit{Networks models can be interpreted as shorthands for joint probability distributions.}

\vspace{0.1in}

Activation values for each node depend on those of many others, rendering graphical models particularly convenient representations of their joint activities. Graphs can explicitly encode the statistical dependencies among different activation variables as connection \emph{weights}, with the states of connected nodes given not by a stochastic transfer function, but by conditional probabilities.

A type of directed acyclic graph (DAG) known as a Bayesian network is a weighted construction whose connection strengths are typically learned~\cite{friedman2000a} via Bayesian inference (i.~e., computing the posterior probabilities for a set of candidate DAGs, and selecting the member with the highest value, etc.) Undirected variants, which communicate only binary dependency information via the presence or absence of symmetric links are popular in different applications. When activities are assumed to deviate normally from baseline values (an assumption that greatly simplifies the inference process), they are known as Gaussian graphical models~\cite{schafer2005}.

Bayesian networks weights can be scaled so as to represent a proper, normalized probability distribution. Adjusted to match that of the observed data, the weights in such a dependency graph become an explicit encoding of the system's joint statistics. Bayesian networks satisfy a \emph{Markov property}, such that the activity value distribution for a given node depends only on the values of its immediate predecessors (these activities are often discretized as binary variables for mathematical convenience, so the resulting graph neatly keeps track of the probability that a downstream node in the inferred network will be active if its predecessors are active). This directed conditional dependency structural arrangement offers a conceptually accessible and intuitive view of the system, although the presence of directed connections between two nodes does not mean there is a direct physical (i.~e., mechanistic) or causal link between the corresponding species~\cite{hartemink2001}.

One of the most important and unique applications of network inference, this compact representation of probability distributions permits the quantitative prediction for nodal activity values, in both static and dynamic contexts.  Probabilistic graphical models are particularly useful in putting numbers on answers to questions like ``What is the probability of this protein being active, given that a particular stimulant is present?'' or its converse: ``What is the probability of the  stimulant having been present, given that the expression level of this gene is high?''~\cite{needham2006}. We discuss methods for inferring both types of probabilistic graphical models named here, and their limitations (including their ability to infer causality), in Section~\ref{reconstruction_methods}.

\subsection{Reconstructions as a part of the Big Picture}
\label{big_picture}

\textit{Inferred network models can be combined with existing and new methods as one part of a larger repertoire for investigating many facets of living systems.}

\vspace{0.1in}

Reconstructions are increasingly combined with other tools and prior biological knowledge to form integrated frameworks for discovery. Some reverse-engineering approaches attempt to incorporate prior knowledge explicitly into the inference process for individual networks~\cite{werhli2007,geier2007,mukherjee2008,greenfield2013,li2015,ghanbari2015}, including one study which advocates the use of undirected gene networks (gleaned from functional association databases) as \emph{priors} to enhance the inference of mechanistic, causal gene regulatory networks~\cite{Studham2014}.

Other applications use networks to cross-reference, corroborate, or pre-screen evidence for predictions about specific systems. For example, the ``network approach'' to genome-wide association studies (GWAS) and disease gene prioritization is reviewed in~\cite{yu2013}, and the use of networks for the prediction of protein functions (in the general sense, not restricted to physical binding), evolutionary studies of pathogenic and non-pathogenic strains, and the bidirectional interactions between host and pathogen are reviewed for the specific context of infectious disease in~\cite{mulder2014}.

We have already mentioned the work~\cite{yamane2016}, which uses Bayesian networks in tandem with support vector machines to predict the toxicity of various chemicals in a supervised setting. Yet we believe the most pivotal roles to be played by reconstructed networks are those which completely change the way we think about biological phenomena, specifically by offering new ways to predict system-wide behaviors. Such a revolution is already underway in medicine: the treatment of various diseases is no longer unilaterally viewed from within the ``one-gene, one-drug'' paradigm, and it is gradually becoming the new standard to view related autoimmune disorders as emanating from a network of maladies with the same root causes~\cite{goh2007,loscalzo2011,zhou2014}.

\section{Two different meanings of phenomenological ``reconstruction''}
\label{reconstruction_methods}

We distinguish two principal categories for phenomenological network inference, accounting for methods that produce undirected and directed graphical models.

Algorithms in our {\bf first category} define an inferred interaction as an {\em irreducible statistical dependency} among nodes, typically quantified by some measure of the similarity among the activation profiles of different nodes. This is a structure-only approach, and should be used when it is only necessary to reconstruct the overall network topology -- in other words, for applications for which it is sufficient to know ``who talks to whom.'' In some cases, topological maps can be augmented with weights that ascribe an effective strength or confidence level to the inferred interactions~\cite{horvath2011,colwell2014}.

Algorithms in our {\bf second category} define interactions in terms of {\em asymmetric relations} capable of describing not only which nodes participate in an interaction, but also ``who controls whom.'' Previous classification schemes have considered the inference of unweighted, directed links as a separate endeavor from discovering quantitative input-output relationships between nodal activities~\cite{rice2005}, or further distinguish algorithms that detect the sign of interactions without an explicit direction~\cite{zhang2005,khosravi2015}. However, since both the types of data and the processing techniques needed to infer all these kinds of graphs are similar, we treat them on equal footing.

\subsection{Who talks to whom?
\emph{Presence, absence of undirected links}}
\label{who_talks}

The most basic question that one can answer in the course of network reconstruction is whether a given subset of nodes can be characterized as interacting -- in other words, \emph{who talks to whom}? Since our focus here is on the unsupervised inference of interaction networks directly from activation data, any notion of ``interaction'' that we consider must depend on these activations alone. A natural definition for the existence of an interaction among species is the presence of statistically significant correlations among their respective activation states. Such a choice results in an undirected network with symmetric (though possibly weighted) connections.

In practice, pairwise statistical dependencies are typically quantified by introducing a \emph{similarity} metric, such as the first-order Pearson correlation. The Pearson correlation coefficient is a normalized, pairwise dependency measure bounded by the interval $[-1,1]$.  Positive (negative) values indicate an increasing (decreasing) linear relationship. While its value is always zero for statistically independent variables, a vanishing Pearson correlation cannot rule out nonlinear correlations. Conversely, in the absence of nonlinear effects, finite sampling can cause independent variables to appear correlated, so that connections can be inferred where no otherwise discernible interaction exists. To avoid inferring such spurious interactions, one must apply a threshold to filter raw correlation values.

When nonlinear effects cannot be ignored, one can quantify statistical dependencies using information-theoretic measures~\cite{shannon1948,cover2012,levchenko2014}, which generalize the notion of correlation to such nonlinear cases. D'Haeseleer et al.~\cite{dhaeseleer1998} were the first to employ the mutual information to uncover gene-to-gene dependencies, while Butte et al.\ applied mutual information ``relevance networks''~\cite{butte2000} to propose single-gene determinants of anticancer agent susceptibility~\cite{butte2000_PNAS} for experimental verification.
Mutual information-based methods must still contend with the same sampling and bias problems faced by linear correlation coefficients, and therefore require thresholding as well.

Even under conditions of perfect sampling, neither Pearson correlations nor the mutual information can disambiguate so-called \emph{direct} interactions from \emph{indirect} interactions -- statistical dependencies that are already accounted for by links involving other species. Note that this notion of ``indirect'' is distinct from its usage in the context of mechanistic networks. There, ``direct'' typically refers to physical contact, which often occurs between nodes whose activations are not included in the network model (unobserved, latent, or marginalized degrees of freedom in the system). Here instead we are concerned with statistical redundancies within the set of \emph{observed} activation variables. For example, consider the case of three genes in a regulatory cascade: $X \rightarrow Y \rightarrow Z$. Inference methods based on measuring correlations between the associated activation variables would find a link between $X$ and $Z$, which is {\em indirect}, in the sense that it is not actually needed to account for the joint statistics of $X$, $Y$, and $Z$.

While sometimes inconvenient, indirect links are not always superfluous. They are useful when probing the network at the single-node level, as when trying to discover a previously unknown member in an established pathway, propose a novel interaction for experimental verification, or predict the overall effect on the activation state of one node by perturbing another. On the other hand, in applications for which inferred networks must be treated as whole entities (e.~g., when they encode normalized probability distributions; see MaxEnt methods described below), this sort of redundancy can be minimized by examining \emph{conditional} dependency structures.

There exist several approaches to studying conditional dependencies. The most intuitive is to work explicitly with either partial correlation coefficients~\cite{delaFuente2004} or the conditional mutual information~\cite{wangK2006,liang2008,wangK2009,margolin2010multivariate,zhang2012} between two activation variables $X$ and $Y$, given another variable (or set of variables) $Z$:
\begin{equation}
I(X;Y|Z)=I(X;Y,Z)-I(X;Z),
\label{eqn:conditionalMI}
\end{equation}
where $I(X,Z)$ is the mutual information between $X$ and $Z$. In principle, one can refine a reconstruction by removing links between any pair of species $X$ and $Y$ that are associated with statistically insignificant values of $I(X;Y|Z)$. However, reliable estimation of this quantity is much more difficult than it is for the pairwise quantities, such as $I(X,Y)$, since it requires sufficiently dense concurrent sampling of at least three variables.

In order to dispose of indirect links without incurring the aforementioned estimation problems, some algorithms make additional assumptions and thus append ancillary filtering steps to the basic mutual information-based procedure. For instance, the Algorithm for the Reconstruction of Accurate Cellular Networks (ARACNe)~\cite{basso2005,margolin2006} invokes the Data Processing Inequality~\cite{cover2012} to delete the weakest link in every closed triplet of nodes (this would be an exact step if the studied network was a tree). The Context Likelihood of Relatedness (CLR) method~\cite{faith2007} determines the presence or absence of a link by assessing its strength against all other mutual information scores computed for that graph, as a background significance threshold. MRNET~\cite{meyer2007} builds a network iteratively, including a link between two variables if one is both a good predictor of the other and yields information that is non-redundant with that provided by the previously inferred links. 

An alternative approach to solving the conditional independence problem is to use full probabilistic models that allow conditioning on the complete set of marginals, rather than requiring the progressive computation of higher-order partial correlations~\cite{zhang2012}. In particular, if a set of continuous, real-valued activation variables are (assumed to be) normally distributed, one can condition a single interaction on the full set of remaining variables. In this case the statistical independence of any two nodes can be ascertained by examining the elements of the inverse of the covariance matrix: $\Sigma^{-1}_{ij}=0$ if and only if $i$ and $j$ are conditionally independent, given all other variables. An important facet of such multivariate Gaussian distributions is that they correspond to the least constrained, \emph{maximum-entropy} models that satisfy the full set of first and second-order marginals for continuous variables~\cite{Jaynes1982,stein2015}. These first two moments correspond to the individual means and the pairwise correlations, which are usually well measured even in sparsely sampled data sets.

Beyond Gaussian variables, the Maximum Entropy principle has been a successful modeling approach in neuroscience~\cite{schneidman2006, ohiorhenuan2010, field2010, tkacik2014}, natural images~\cite{bethge2007}, the inference of gene networks (from expression data)~\cite{lezon2006} and signal transduction networks (from phosphorylation proteomics data)~\cite{locasale2009}, and the prediction of amino acid contacts in proteins~\cite{weigt2009,morcos2011,marks2011,jones2012}, multidrug effects~\cite{wood2012}, protein structural attributes~\cite{halabi2009}, antibody diversity~\cite{mora2010}, and even the dynamics of flocking birds~\cite{bialek2012}. The joint probability distribution for a maximum entropy model has a particular form, known in statistical mechanics as the Boltzmann distribution. If we ask to match only the empirical means $\langle x_i \rangle$ and pairwise correlations $\langle x_i x_j\rangle$ to those of the observed data, the distribution with maximal entropy is

\begin{equation}
\displaystyle P(\vec{x}) = \frac{1}{\mathcal{Z}} \exp\left( \sum_i h_i x_i+ \sum_{ij} J_{ij}x_i x_j \right).
\label{pairwise}
\end{equation}
Here parameters $h_i$ and $J_{ij}$ are known as the fields and the couplings, respectively, and $\mathcal{Z}$ is the partition function (compare to the full expansion in Section~\ref{goals}).

For discrete variables, the Maximum Entropy model retains the form of Eq.~\eqref{pairwise}, but is known as the Ising model (for binary variables) or Potts model (for categorical variables with more than two accessible states). In the discrete case, fitting the parameters $\left\lbrace h_i,J_{ij} \right\rbrace$ is highly nontrivial. Many methods exist, but their effectiveness depends on the system size and the density of its interactions, as well as on other properties~\cite{ackley1985, roudi2009, hastie2009, ravikumar2010, cocco2011}. One algorithm worth mentioning is the \emph{adaptive cluster expansion}, which was developed in the context of the MaxEnt problem~\cite{cocco2011, cocco2012}. It is closely related to information-theoretic approaches, being equivalent to relevance networks~\cite{butte2000} for clusters of size two, and similar to conditional mutual information methods for clusters of size three.

Due to the limitations of finite sampling, both solving for the inverse of the covariance matrix and learning the parameters of an Ising model can constitute ill-posed problems. One way to avoid this is to impose a regularization~\cite{bishop2006}, which invokes additional constraints on the interaction coefficients to ensure that the inference problem is well-defined -- and moreover, that the inferred network generalizes well to unseen data. Regularization is often done in one of two common ways: either the interactions coefficients are assumed to be small (for example, using an $L_2$ norm)~\cite{cocco2012} or the interaction structure of the system is presumed to be sparse, so that the overall number of the interactions is small (this may be done explicitly by specifying the number of non-zero coefficients \cite{gardner2003} or by invoking an $L_1$ norm~\cite{ravikumar2010}).

Frequently cited as the rationale behind these regularization procedures is the inherent sparsity of natural networks~\cite{jeong2000,jeong2001,leclerc2008}. Indeed, for protein studies, the nodes in networks used to describe tertiary protein structure represent real amino acids in the three-dimensional space; they can therefore be connected to only a small subset of all possible neighbors. Similarly, the number of transcription factors that can influence a given gene's expression levels is limited by the physical extent and arrangement of its promoter sequence. While the general ubiquity of sparseness in biological systems is debated~\cite{delaFuente2002}, the enforcement of sparsity constraints can be justified as a purely pragmatic measure in the ``low-hanging fruit'' inference regime.

\subsection{Who controls whom? \emph{Causal relations and directed links}}
\label{who_controls}

Directed network inference differs in an important way from that of undirected, symmetric, mutual-influence graphs: since questions of causality (or, more generally, the  flow of information) are built not upon a single, universally agreeable concept like statistical correlation -- but rather on more subtle, less straightforward notions like \emph{control} -- there exist many diverse criteria for establishing directed connections. Each method has its own operating definition of what counts as an interaction, and how to infer its direction.

Though disparate, the aforementioned definitions can be conveniently divided into approximately two subclasses, depending on the intended application of the inference procedure. In certain cases, it is enough to know the direction or causal \emph{sense} of an inferred interaction. For example, will silencing a certain gene or disabling a particular neuron result in a collapse of the entire system? Can the intracellular concentration of a reactant be increased by introducing more of the product? Answers to questions like these do not require numbers, entailing purely qualitative predictions. On the other hand, if the goal is to use a reconstructed network to predict the amount by which one gene's expression level increases when two other genes are suppressed, directed connections must be weighted by quantitative values representing the effective \emph{strengths} of interactions. We describe methods of both types, leaving it as an exercise for the reader to think about when a directed topology suffices, and when it is necessary to infer fully signed and weighted graphs.

Before we delve into specific methods, we advise the reader to tread with caution. The particular definitions of directed influence we explore in the following methods do not always correspond to our intuitive and/or mathematically formal notions of causality. As a result, producing a graph with directed links does not automatically satisfy a reverse-engineer's desire to uncover system-wide causes in an ontological sense, and should not be treated as such despite one's instincts. Instead, great care needs to be taken with each method in order to ensure that all idiosyncratic constraints are met, and to avoid generalizing or extrapolating beyond the predictive power of each algorithm.

To expound on this point, it is worth asking at the outset whether it is even possible to infer causal information from passive observations of activation variables~\cite{karvanen2014estimating,nitzan2017revealing}. It has long been  understood~\cite{pearl2009} that proximal causal relations can be inferred reliably when the observer is able to \emph{interact} with the system in accordance with a principled protocol (as is done in many controlled experimental interventions~\cite{wagner2001,molinelli2013}, including genetic knockouts~\cite{pinna2010,ud2016} and multifactorial perturbations~\cite{jansen2003,tegner2007}). While this is old news to engineering audiences, it has also been shown that causal information (or at least a lower-bound estimate of causal effects) can be extracted from purely observational data when the equivalence class for the fully directed graph can be ascertained first~\cite{maathuis2009,maathuis2010}\footnote{Once the equivalence class is determined, formal causality detection methods can be subsequently applied to estimate the full causal graph. We refer curious readers to~\cite{pearl2009,spirtes2012} for a wealth of both philosophical explications and more rigorous treatments of algorithms designed explicitly to detect causality in its many guises.}.

We mention again a surprising corollary of this result that directed influence (a less stringent condition, and slightly less nebulous concept, than causal influence) can often be established without time series data, using only static measurements. Where there was once a prevalent belief in the reverse-engineering community that the inference of directed edges required temporal data~\cite{margolin2006}, there is now a tradition of algorithms which accept static data as inputs ~\cite{wagner2001,delaFuente2002,yeung2002,gardner2003,Tegner2003,sontag2004,djordjevic2014}. However, we focus for coherence predominantly on methods that operate on time-series data. 

\vspace{0.1in}

We organize this subsection as follows. We first make a few general remarks about the inference of directed interaction patterns. We then explore a class of methods which presume that the measured activities can be treated as deterministic variables that change smoothly in accordance with a particular, predetermined quantitative law. Afterwards we switch to model-free deterministic methods, for which there is no need to specify a mathematical form or law in advance in order to detect interactions. We then treat the more general situation, in which activations are regarded as stochastic variables. Again we start with methods requiring a parameterized model and conclude with a discussion of stochastic model-free methods.

\vspace{0.1in}

A na{\"i}ve but conceptually intuitive approach to inferring directed connections is to take the presence of strong temporal correlations between the trajectories of different activity variables as evidence for a (casual) interaction between the corresponding species. It is common for changes in one activity variable to succeed that of another in time (consider a gene whose expression level is observed to increase consistently in response to the elevation of another), but the proxy of temporal precedence is not robust as a criterion for declaring control relations~\cite{sugihara2012} because it also appears in the absence of causal influence. Despite its limitations, this strategy, combined with a projection method known as multidimensional scaling~\cite{arkin1995} in an algorithm entitled ``Correlation Metric Construction,'' was originally used to infer the first steps of the glycolytic pathway~\cite{arkin1997} and more recently applied to study the pharmokinetics of the anticancer drug Gemcitabine~\cite{lecca2012}.

In physics and engineering, signed and directed connections are often used to encode the weighted coupling constants that appear in systems of differential equations~\cite{ljung1999}. To write down such a system, one needs to first have in mind a particular quantitative form for a dynamical law, according to which activations will be presumed to interact. One then fits the model parameters, typically with some optimization or statistical learning technique that takes time series data as input, and reports the learned values as the weights for the corresponding connections, sometimes adding additional, unobserved, hidden variables in the process \cite{daniels2015}.

The inherent directionality of this method, which works best for small systems ($p\sim 10$), can be understood immediately by examining the matrix $J_{ij}$ of pairwise interactions in Eq.~\eqref{eqn:dynamicalsystem} below: since this matrix is not constrained to be symmetric, couplings between two species can differ in the forward and backward directions. For continuous activation variables $\left\lbrace x_i\right\rbrace$, many popular models can be subsumed as special cases of the general form (though see \cite{daniels2015,schmidt2009distilling} for alternative forms):
\begin{equation}
\frac{d x_i(t)}{dt}=f_i\left(x_i + \sum_j^p J_{ij}x_j + u_i+\xi_i\right),
\label{eqn:dynamicalsystem}
\end{equation}
which includes at most pairwise interactions of strengths $\left\lbrace J_{ij}\right\rbrace$ between all element pairs $i$ and $j$. Here the functions $\left\lbrace f_i\right\rbrace$ can be chosen according to the desired level of computational complexity (controlled by the amount of data available) or biochemical detail, or both. In the reverse-engineering of biological networks, many early applications were linear activation models~\cite{chen1999,dhaeseleer1999,weaver1999,mjolsness2000}, for which $f_i(x)\propto x$. The sum determines the net (excitatory and inhibitory) effect on the activation of node $i$ at time $t$, given its interactions with all other elements $j$. The next term accounts for external driving of the node, (i.~e., any extrinsic perturbation that increases or decreases its activation value by an amount $u_i(t)$), and $\xi_i(t)$ represents noise.

Linear, ``additive'' regulatory models are based on the assumption that dynamical systems can be \emph{linearized} about their steady-states. They are relatively easy to fit in sparsely sampled conditions, especially when the terms in Eq.~\eqref{eqn:dynamicalsystem} are discretized to form a linear difference equation~\cite{weaver1999,vanSomeren2000}. Early work countered undersampling by augmenting the number of data points for multilinear regression via nonlinear interpolation~\cite{dhaeseleer1999}, or imposing sparsity constraints on singular decomposition algorithms~\cite{yeung2002}. Another approach to decreasing the number of interactions that must be inferred is to first cluster the nodes~\cite{bonneau2006}. In any case, data are typically taken during the system's approach to steady-state conditions (whether its natural equilibrium or another fixed point of its dynamics) after a perturbation.

A straightforward modification of the basic linear model, realized by overlaying the sum in Eq.~\eqref{eqn:dynamicalsystem} with a sigmoidal threshold function, leads to one version of the \emph{artificial neural network} construction. Early methods based on neural networks were used to infer interactions between individual~\cite{Wahde2000} and aggregate ``genes'' which encompass multiple degrees of freedom at the biological level~\cite{mjolsness2000}. Modern improvements use multilayer perceptrons~\cite{grimaldi2011}. Early neural-inspired architectures known as gene circuits~\cite{mjolsness1991} have also been used to infer mechanistic interactions~\cite{crombach2012}.

Nonlinear models are attractive because they can capture more sophisticated dynamical behaviors than their linear counterparts (e.~g., oscillations and multistability). Nonlinear reverse-engineering schemes based on mass-action kinetic laws like Michaelis-Menten or Hill equations~\cite{alon2006} are also used in reconstruction~\cite{oates2014,Mangan2016}.

\begin{figure*}[th!]
\centering
\includegraphics[width=5in,angle=0]{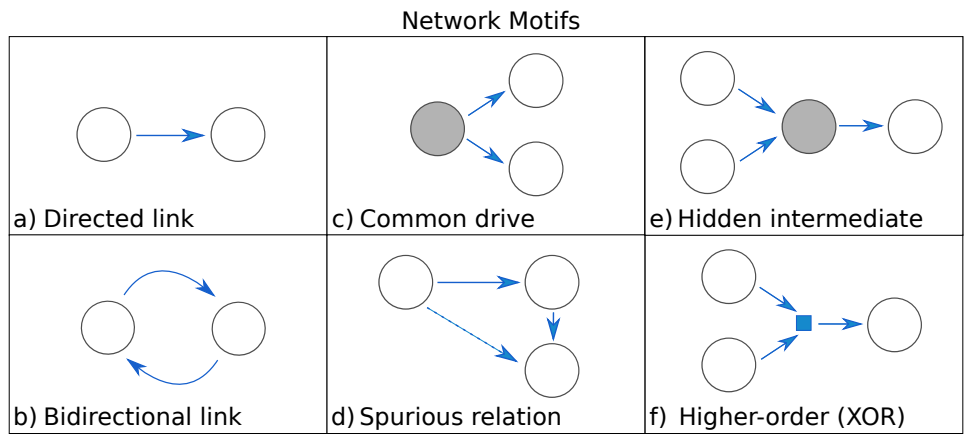}
\caption{\small \label{fig:motifs}{Simple directed network motifs help illustrate basic problems in directed network reconstruction. This list is not intended to be comprehensive, but to address some simple yet important scenarios. Links can represent mechanistic interactions or effective relations (i.~e., information transfer). Nodes represent stochastic or deterministic activation variables, which can be either continuous or discrete. Here, dashed links represent spurious (erroneously inferred) interactions, dark nodes represent unobserved (hidden) variables, and the small square in f) refers to a computation that involves more than two nodes (in this case, a third-order interaction). \textbf{a)} The simplest scenario: a directed link between two nodes. \textbf{b)} A bidirectional coupling models a simple system with feedback (e.~g., the predator-prey system of Fig.~\ref{fig:figure_1}). \textbf{c)} A hidden common drive (dark node) to two observed nodes results in a correlative relation between those nodes. If care is not taken, this can be confounded with a direct causal interaction. \textbf{d)} A situation similar to that of c), with the difference that measurements of all three nodes are accessible. Na{\"i}ve pairwise methods infer a spurious link between the initial and final nodes in the feedforward chain. Multivariate methods are required in this scenario to infer the correct links. \textbf{e)} In the case of a hidden node relaying the causal interaction, network reconstruction methods may infer the correct direction of interaction, but the inferred links will be effective rather than strictly causative since an intervention at the hidden node can disrupt the interaction. 
\textbf{f)} The logic gate XOR entails a higher-order interaction. The output is 0 if both input nodes carry the same value, and 1 if they are different: simultaneous knowledge of the states of both nodes is required to determine how each of the inputs affects the output. This is a classical example of a scenario where X and Y carry synergistic (as opposed to unique, or redundant) information~\cite{wibral2015}.}}
\end{figure*}

An important causal inference method based on the assumption of an underlying deterministic system, but which does not require the definition of an explicit dynamical model, is the convergent cross-mapping (CCM) approach \cite{sugihara2012}. As noted in~\cite{luo2015}, an essentially identical method had been developed earlier to study synchronization in chaotic dynamical systems~\cite{rulkov1995}. The method draws from Takens' theorems~\cite{takens1981}, which provide both the conceptual framework and mathematical justification for a brand of state space reconstruction -- reverse-engineering of the phase-space portrait for a dynamical system -- known as \emph{delay embedding}. Consider a multidimensional dynamical system, a special case of the general form~\eqref{eqn:dynamicalsystem} whose parameters are fixed, and whose temporal evolution $\mathbf{x}(t)$ is confined to a subspace determined by a $d$-dimensional attractor~\cite{strogatz2015}. Under very general conditions, the attractor's state space can be reconstructed ~\cite{takens1981} from measurements of a single time series $\{x_{t},x_{t+\tau},x_{t+2\tau},\ldots\}$, sampled at an interval $\tau$. The number of consecutive time points needed to span the reconstruction space is given by the attractor dimension $d$; both $\tau$ and $d$ are often found using Ragwitz' criterion~\cite{ragwitz2002}, but alternative methods have been proposed as well~\cite{small2004, faes2012}.

Delay embedding refers to the entire process of defining these two parameters and arriving at a reconstruction space onto which the time series can be mapped. It provides the substrate for causal inference via CCM as follows. For any two measured times series $\{x_t\}$ and $\{y_t\}$, the variables $x$ and $y$ are said to be causally linked if they belong to the same underlying dynamic system (i.~e., the time series they represent are samples from the same attractor~\cite{takens1981,sugihara2012,strogatz2015}). The direction of an interaction between $x$ and $y$ variables can be estimated by 1) using delay embedding to obtain reconstruction manifolds $\mathcal{M}_x$ and $\mathcal{M}_y$ for $x_t$ and $y_t$, respectively~\cite{strogatz2015}; 2) projecting one of the variables, say $x$, onto the other manifold -- hence the name cross-mapping -- and using the resulting, projected values to predict the values taken by the original time series (which \emph{converge} to the measured values for a large enough number of samples); and 3) measuring (with any suitable measure, e.~g. RMSE or correlation function) the deviation of the postdicted values $\{\hat{x}_t\}$ from the actual values $\{x_t\}$. A causal interaction is declared if the prediction quality does not decay to zero for a growing number of samples.

In the original work, Sugihara et al.~\cite{sugihara2012} did not analyze thoroughly the influence of noise on reconstruction. Indeed, Takens' original theorems allow for noise in the measurement procedure only (i.~e., intrinsic stochasticity is prohibited; the breakdown of inference based on CCM in the presence of intrinsic noise has been demonstrated explicitly~\cite{mccracken2014,cobey2016, monster2016}, and a thorough analysis of state space reconstruction in the presence of noise can be found in~\cite{ragwitz2002}). Nevertheless, artificially added measurement noise can actually improve the detection of causality~\cite{jiang2016}.

Several other considerations must be taken into account when inferring causal relations with CCM. First, it seems that the outcome is quite sensitive to the sampling methods used to obtain training data (for example, eliminating nonstationarity on the way to the attractor is key)~\cite{luo2015}. Second, CCM fails to infer the accurate coupling strengths and even the direction of causal interaction when time series are synchronous~\cite{monster2016}. Third, it has been shown that the predictions made by CCM do not always conform to our intuitive notions of causality, even for certain rudimentary systems like a simple resistor-inductor (R-L) circuit with a sinusoidal driving voltage, where CCM does not unequivocally determine the causal dependence of the current on the voltage~\cite{mccracken2014}. Finally, Cobey and Baskerville~\cite{cobey2016} provide a thorough numerical analysis of the limits of CCM, suggesting that the standard approach is generally prone to failure if the system dynamics are oscillatory and proposing a modification in the algorithm to alleviate this shortcoming~\cite{cobey2016}.

For stochastic activations, early attempts to reconstruct the directionality of interactions included autoregressive models~\cite{wiggins2003process,opgen2007learning}, but autoregression by itself makes no assertions about causality. However, a method due to Granger~\cite{granger1969} combines autoregression with the aforementioned notion of temporal precedence to infer quantify a robust stand-in for causality -- namely, Weiner’s predictability~\cite{wiener1956}. The framework for Granger Causality (GC) is built upon two central assumptions~\cite{granger1988}:
\begin{enumerate}
\item The cause $x$ occurs before the effect $y$.
\item The causal series $\{x_t\}$ contains unique information about the time series being caused $\{y_t\}$ that is not available in any other series $\{w_t\}$.
\end{enumerate}
More generally, $\{w_t\}$ represents the entirety of processes that can influence $\{x_t\}$ and $\{y_t\}$. In the ideal scenario, for which these three variables together contain ``all the information available in the universe at time $t$"~\cite{granger1988} (i.~e., in the closed system under investigation), GC guarantees that one can reconstruct the direction of the causal relationship between $x$ and $y$. By definition, a variable $x$ ``Granger-causes'' variable $y$ if knowledge of past values of both $x$ and $y$ reduces the variance of the prediction error for $y$, in comparison with the history of $y$ alone. Typically, these predictions are carried out via linear regression, and the direction of causality is decided by statistical tests on the variances of the respective residuals (prediction errors). However, this implicitly assumes (at most) linear relations between variables. Nonlinear extensions of GC exist, but these extensions can be more difficult to use in practice and their statistical properties are less well understood~\cite{freiwald1999,ancona2004,chen2004,gourevitch2006,marinazzo2008}.

Granger causality can be extended to multivariate scenarios~\cite{ding2006} as well, although finding Granger-causal links among all possible candidate interactions then becomes a combinatorially hard problem. For the particular case of inferring causal relations between the activity of distinct brain areas (using electroencephalograms or local field potential time series), it has been found to be of crucial importance to employ a multivariate approach rather than bivariate techniques~\cite{blinowska2004}.

A more general approach to the reverse-engineering of directed links between stochastic variables is to learn an explicit model for the joint probability distribution of the observed activities. This approach, based on \emph{probabilistic graphical models}, was discussed earlier for undirected networks. For the directed case, one can define a class of models known as Bayesian networks~\cite{pearl1985,charniak1991,ghahramani2001,friedman2004} which decompose the joint distribution into separate factors representing conditional probabilities. Edges are drawn starting from the nodes corresponding to variables being conditioned on (called the ``parents'') and ending on the conditioned variables (the ``children'')~\cite{friedman2004, pearl2009}. Since the joint distribution of a Bayesian network is an exact product of conditional probabilities, the resulting graphical structure is a \emph{directed acyclic graph} (DAG). Thus in order to be eligible for representation by a Bayesian network, systems need to satisfy the necessary criteria for forming a DAG. If the phenomenon in question is known to encompass cyclic dependencies (e.~g., autoregulation pathways in gene regulatory networks, or autapses in neural networks), the only recourse is to ``unroll'' the cyclic dynamics in time, forming a \emph{dynamic Bayesian network}~\cite{friedman1998,hartemink2001,smith2002,nachman2004,acerbi2014}. The performance of dynamic Bayesian nets has been been compared directly against that of Granger causality~\cite{zou2009granger}, and favorably so when the observed time series are shorter than a certain length (NB: In general, findings like these should be taken with a grain of salt, since 1) they could be artifactual results that depend on idiosyncratic features of the data, and 2) notions of error and accuracy tend to rest on the existence of a reference network containing only the ``correct'' edges, which is in our opinion a dubious concept; see comments on evaluation metrics in the \emph{Discussion}. In~\cite{zou2009granger}, the authors are clear in their admission that ``the causal relationship derived from these two approaches could be different, in particular when we face the data obtained from experiments,'' in accordance with our introductory statements about the nonuniform definitions of causality that are assumed by different methods.).

With the conditional probability framework in place, one needs to select 1) a quantitative form for the underlying model that parameterizes the conditional probabilities, 2) a scoring or objective function that quantifies the quality of fit, and 3) an optimization or search routine by which to learn the parameters values that extremize the objective function. An example of such a parameterization, used quite frequently in the literature, is again that of linear regression~\cite{friedman2000a,friedman2004}. The choice of a specific parametric representation of conditional probabilities is often dictated by our knowledge or assumptions about the domain (prior knowledge)~\cite{zhu2008}, or pragmatic principles favoring computationally simple models (Occam's razor). Standard objectives are the maximization of the likelihood function~\cite{friedman1998} or posterior probability distribution~\cite{friedman2000a}, as well as the Bayesian Information Criterion (BIC)~\cite{nachman2004}, which penalizes for large numbers of parameters. Since the optimization search is an NP-hard problem~\cite{charniak1991, friedman2004}, exact methods are often computationally infeasible, so one often reverts to heuristics like greedy hill-climbing (which adds, deletes, or reverses edges to encourage maximal ascent in the objective score~\cite{pournara2004}), stochastic hill-climbing, or Monte Carlo methods~\cite{friedman2000b}.

An impressively comprehensive and thorough body of work regarding the concept of causality and its formal description via Bayesian nets has been provided by their originator, Judea Pearl~\cite{pearl2009}. Pearl introduces a conceptual framework called the \emph{do-formalism} (known variously as the do-calculus, the intervention-calculus, etc.), which formally describes the use of experimental interventions to ascertain a causal structure. In the do-formalism, $p(y|do(x))$ denotes conditioning on a variable $x$ that is experimentally controlled rather than simply measured (i.~e., observed passively). In other words, this notation distinguishes the more familiar observational conditioning $p(y|x)$ from ``interventional conditioning''~\cite{ay2008,emmert2013}.

While correlation does not in general imply causal influence, Pearl reveals specific cases for which the conditional probability distribution -- reflecting associative dependencies -- is equivalent to that which denotes the corresponding mechanistic dependencies: in such situations, interventions which manipulate the values of parent nodes are clearly and unambiguously seen to have direct effects on the children, and the Bayesian graph is therefore also the correct casual graph. 

It is often difficult to satisfy all the criteria for modeling a causal system with DAGs. In certain circumstances, it is easier to work with model-free stochastic frameworks, such as that of the transfer entropy (TE). TE was introduced twice independently, by the physicists Schreiber~\cite{schreiber2000} and Palu\v{s}~\cite{paluvs2001}, and has since proven to be a versatile and useful tool for inferring the direction of information transfer in neuroscience~\cite{honey2007, stetter2012, wibral2014}, physiology~\cite{faes2012, faes2013}, climatology~\cite{pompe2011, runge2012a, runge2012b} and economics~\cite{kwon2008, kim2013}.
TE is simply the conditional mutual information~\eqref{eqn:conditionalMI} between a target variable $Y$ and the entire history of values assumed by a source variable $X$, given the history of the target:
\begin{equation}
\mathcal{T}(X \rightarrow Y) = I(\textbf{X}_{t-};Y_t|\textbf{Y}_{t-}).
\label{eqn:TE}
\end{equation}
Here the arrow denotes the direction of information transfer (i.~e., $X$ informs $Y$) and $\textbf{X}_{t-}$ and $\textbf{Y}_{t-}$ respectively denote the histories of the corresponding stochastic processes up to, but not including, $t$; $Y_t$ denotes the value taken by the target variable at time $t$. Conditioning on the history of the target ensures that only those bits of information that are unique (in the sense discussed earlier for Granger Causality; for a formal treatment see~\cite{bertschinger2014, wibral2015}) to the source variable are considered.

Like all information-theoretic measures, TE and its surrogates~\cite{sugihara2012} suffer from the curse of dimensionality because of the need to estimate entire probability distributions (discrete variables) or probability densities (continuous variables) for long time series and many variables. For discrete variables, the simplest estimation procedure entails simply counting frequencies to produce a histogram that approximates the desired distribution. A substantially more accurate estimation of information-theoretic quantities for discrete variables (especially if the data set is small) can be obtained by computing entropies directly with the NSB estimator~\cite{Nemenman2002,nemenman2011}. In the continuous case, a standard approach is to bin the data, rendering the distribution effectively discrete and therefore amenable to histogram methods. While less ``data hungry'' alternatives exist for continuous variables (such as kernel estimators~\cite{friedman2001}), they suffer from the same systematic estimation biases that are associated with histogram methods~\cite{kraskov2004}, and may even reverse the inferred direction of information flow~\cite{hahs2011}. Nearest neighbor estimators~\cite{kraskov2004, wibral2014} are some of the most commonly used in practice. In all cases, statistical testing against surrogate data or empirical control data~\cite{lindner2011} is recommended to help ameliorate the bias problem.

An approach to dimensionality reduction based on the concept of Markov chains has been proposed for the estimation of TE~\cite{runge2012a}. This approach is particularly useful in the case of delayed coupling between variables~\cite{wibral2013}:
estimation of the delay time can prevent the inclusion of unnecessary time steps when tracking the history of the source variable (i.~e., $X_{t-}$ in Eq. \eqref{eqn:TE}), which can clearly reduce the dimensionality of the latent representation. Finally, the curse of dimensionality can also be alleviated by first constructing an explicit, low-dimensional model of the time series (and hence, parameterizing the probability distribution). For the simplest case -- linear dependence between $X$ and $Y$ with additive Gaussian noise -- it has been shown analytically that TE will always recover the same network as Granger Causality, up to a constant factor~\cite{barnett2009}.

Since some authors speak loosely about inferring causality when computing the TE or related quantities like the directed information~\cite{kaleta2010}, we reiterate that, although causal interaction is a necessity for information transfer, the converse is not true: information transfer, as quantified by TE and other information-theoretic functionals, does not imply underlying causal interactions. In fact, we caution readers that some methods for the detection of causal or directed influence have been routinely applied in ways that differ markedly from the intentions of their originators. For instance, the directed information was initially designed to infer achievable information rates on a known communication channel with feedback~\cite{massey1990}, rather than the inference of directed networks (for a thorough discussion, see~\cite{wibral2014}). However, TE specifically has been extended using the aforementioned do-formalism in a new procedure known as \textit{information flow}~\cite{pearl2009}, a more appropriate measure for inferring causality under certain constraints~\cite{ay2008, lizier2010}. Notably, this measure can correctly resolve the connectivities of an XOR circuit (see Fig.~\ref{fig:motifs}\textbf{f)}) even in special scenarios where the conditional mutual information fails~\cite{ay2008}, a fact overlooked by authors who have contended that conditional mutual information is sufficient for this purpose (see, for instance, the argument in ~\cite{liang2008}). Finally, we note that TE and similar methods have not achieved widespread implementation for large systems ($p \gg 1$) due to the aforementioned, intrinsic difficulty of estimating information theoretic measures in high dimensional spaces. Multivariate approaches to TE estimation and related methods are a subject of ongoing research.

\section{Discussion}

Since the year 2000, some thirty review articles that we know of have been published on the inference of gene networks alone (in addition to those referenced or mentioned throughout, see~\cite{dhaeseleer2000,wessels2001,Brazhnik2002,deJong2002,vanSomeren2002,albert2004boolean,vanRiel2006,albert2007,choKH2007,Goutsias2007,Markowetz2007,Kaderali2008,Karlebach2008,Bonneau2008,lee2009computational,shmulevich2009,sima2009,ay2011,penfold2011,veiga2010,hernandez2013,ristevski2013,vijesh2013,villaverde2013,Emmert-Streib2014,wangYXR2014,dong2015,Liseron-Monfils2015}), and an increasing number have begun to specialize on the unique challenges faced by network reverse-engineers rather than merely listing different algorithms~\cite{marbach2010,Oates2012,Noor2013,Omony2013,yu2013,thomas2014,linde2015,Aijo2016}. One DREAM report~\cite{stolovitzky2009} notes that the number of PubMed articles on reverse-engineering had doubled each year for over a decade through 2009, and ``novel'' algorithms (new twists on the same foundational principles we outline above) continue to emerge even as we write~\cite{budden2016}.

Has this explosive growth in the number of reverse-engineering algorithms and studies helped carve out a niche for large-scale reverse-engineering in contemporary systems biology repertoires? Or has a staunch directive on the reconstruction of entire microscopic networks actually encumbered and obfuscated our understanding of the working principles that underlie these complex systems?

One major impediment to assessing the promise of reverse-engineering algorithms stems from the way in which they are assessed: we observe a rampant, pervasive, and potentially counterproductive tendency to draw direct, quantitative comparisons between reconstructions produced by different algorithms. In other words, despite the commoditization of network inference tools, there is still no consensus on the correct way to \emph{evaluate} reconstruction results~\cite{stolovitzky2007} --  and perhaps for good reason! In the context of effective network inference, the notion that reconstructions can be \emph{checked for accuracy} contradicts our very premise, that algorithms both among and within each of the classes we have described make diverse assumptions about what should count as an interaction. Recent work~\cite{boutros2014,siegenthaler2014} notwithstanding, we believe this issue continues to be confounded by a repeated mismatch between algorithms and metrics (as in the use of the area under receiver-operator characteristic curves~\cite{Hand2009}, a measure that presupposes the existence of a valid confusion matrix, to give an overall rank or ``score'' to effective reconstructions~\cite{Oates2012,Maetschke2013}).

The methods in different classes also differ in more concrete ways: they vary in the extent to which they can infer strengths, signs, and directions for the interactions they detect. This might be thought of as a ``feature, not a bug'' of reverse-engineering technologies: having a selection of versatile algorithms, each tailored to particular situations or designed with different inference goals in mind, increases the chances that researchers can make use of reverse-engineering algorithms. Yet the question of whether systems biologists should persist in pursuing whole-network reconstruction as a go-to modality or learning tool hinges not solely on whether the inference goals are achieved by the time the smoke clears, but on the attainment of a reasonable \emph{tradeoff} between the computational effort consumed by inference algorithms and the (ideally, unique) benefits they afford to researchers.

\vspace{0.1in}

Do the spectrum and short history of network inference successes live up to such high hopes? Along these lines, we have argued that reverse-engineering over the past two decades has played at least five distinct research roles -- the acceleration of hypothesis generation and verification at the single-node/single-interaction level, the illumination of statistical properties that render biological networks unique among complex systems, the diagnosis of individual networks as either typical or perturbed (paralleled by the use of within-class variation to make theoretical statements about the system), the prediction of how the activities in a given network will respond to exogenous perturbations, and the compact encoding of joint probability distributions -- that go far beyond the trivial task of piecing together which of a set of observed system elements engage in physical contacts or the transfer of biologically relevant information. The roles we have identified represent a far cry from the (three) uses of effective influence networks -- identification of functional modules, probing the response to perturbations, and helping determine the underlying mechanistic interactions -- named by the authors of Ref.~\cite{bansal2007} ten years ago.  

While it is impossible to say which of recent attempts to use networks as compressed ``statistics'' to help make (quantitative or qualitative) predictions will have the biggest impact down the road, it is clear that new precedents for the prediction of drugs targets and systemic responses in network medicine~\cite{csermely2013} point to a significant departure from the more traditional, reductionist ways of thinking. The consequences here will almost certainly include dramatic impacts on the ways medicine is practiced in the lifetime of the reader. With this example in mind, we reiterate our assertion that reverse-engineering yields its most succulent fruit when it is used to augment other methods of expanding our understanding of how living systems work, rather than employed disposably as an end goal in itself. Indeed, changes in the ways network inference has been used over time seem to be in accordance with this sentiment: whereas in 2003 the field was still firmly entrenched in its ``pattern-detection phase''~\cite{goodman1999} (to better understand the state of the art at that time, we recommend~\cite{alm2003}), it was around the time of publication of~\cite{stolovitzky2009} in 2009 that the DREAM4 Challenges first introduced predictive modeling tasks as part of the main annual competition.

Indeed, the DREAM competitions play a unique part in the reverse-engineering culture.  They not only echo changes in the field's priorities but also inform them: they have helped set the precedent in establishing inferred networks as tools for making predictions (as in the DREAM8 prompt to anticipate the responses of cellular signals to yet-unseen perturbations~\cite{zhu2014}). More radically, some of the most recent Challenges go as far as skipping the hitherto-canonical intermediate step of network inference entirely, asking competitors to infer macroscopic properties or outcomes using wholly different types of data~\cite{noren2016}. While we clearly do not advocate for the complete abandonment of automated, network-scale reverse-engineering from large data sets, we do view the foundation's decreasing reliance on methods which require the construction of a detailed microscopic model prior to making inference about the macroscopic system as a progressive step. In fact, we contend that, given suitable alternatives, whole-network reverse-engineering may not be justified in every case.

\vspace{0.1in}
If the reverse-engineering of entire microscopic networks is not always the right tool for the job, what might be done instead? As a starting point, we suggest asking:

\begin{itemize}
\item Given a reverse-engineered network, can we find any further compressions of that network that still preserve information about (i.~e., are equally good at predicting) the macroscopic properties and observables it encodes?

\item Can we identify any coarse \emph{functional units} (perhaps with their own set of interaction rules and dynamics) that might supplant individual nodes and edges as the elements of a common parlance for the study of living systems?
\end{itemize}

For instance, might more appropriate ``parts lists'' for biological systems consist not of individual species' activations, but of larger physical or conceptual elements (e.~g., negative feedback loops and operons) with their own dynamical interaction laws? Alternatively, attractors of the dynamics of biological networks may be a more laconic descriptions of the networks than the interactions among the nodes themselves \cite{Lang:2014cr,merchan2016}. This possibility may be motivated via a historical analogy: \emph{renormalization group} theory in physics~\cite{wilson1979} offers a systematic way to deduce an appropriate new vocabulary (and the corresponding syntax) when one changes the physical scale at which a system is to be observed. The effective interaction rules which emerge (say, the interactions between groups of Ising spins) are not always easily reducible to the familiar dynamics of microscopic activation variables (the nearest-neighbor interactions associated with individual spins), but which nonetheless account accurately for their effects at the new scale.

A recent line of work, inspired directly by statistical physics, formalizes the argument that only a small subset of parameter combinations are easily learnable from data, and therefore that only certain (combinations of) microscopic parameters can be \emph{relevant} in determining a complex system's macroscopic or emergent properties~\cite{machta2013,mehta2014,transtrum2015,bradde2016}. By systematically integrating out ``sloppy'' parameters or parameter combinations, whose values remain relatively unconstrained, one can assemble coarse, parsimonious models in terms of the remaining ``stiff'' parameters that serve as effective, low-dimensional compressions of a system's microscopic statistics.

Answers to the second question -- that of finding higher-level explanatory structures in terms of which system's behavior can be understood -- have been explored since the inception of ``module-based'' inference~\cite{maslov2002,milo2002}. In fact, newer and more powerful tools have sparked a resurgence~\cite{Joshi2009,michoel2009,Bonnet2010,deSmet2010} of this approach. Around the same time, it was demonstrated that the flow of information in development, from promoter sequence to expression, can be reliably understood in terms of coarse, multiple-sequence patterns called graph-mers~\cite{li2010} that encompass entire sequence motifs. Ultimately, we believe that it will be work in directions such as these, which involve gross reconceptualizations regarding the fundamental actors in the biological dynamics, that will supersede whole-network reverse-engineering. 

If the end goal of emulating physics-style modeling is prediction, the penultimate is certainly intuition and conceptual understanding. We entertain~\emph{phenomenological} approaches like renormalization because they promise to yield interpretable models, not intractably large sets of detailed equations. Yet we still stress that, while searching for modularity and simple descriptions entails an invocation of the engineering \emph{mindset} that has informed systems biology since its inception, the principles of good biological design often differ markedly from what works in that context; an open mind is necessary to dream up fitting new constructs. Whatever the case, we are confident that it is only by focusing on phenomenological (rather than microscopic) accuracy that we can deliver a satisfying confutational blow to famous Rutherford's quip that ``all sciences are either physics or stamp collecting”~\cite{birks1962} and begin removing the major impediments to the advancement of formal theories in biology~\cite{bialek2012_book}.

\section{Acknowledgments}

This work was funded in part by the James S. McDonnell foundation Grant JSMF/ 220020321 and the National Science Foundation Grant NSF/PoLS/1410978, and by the Laney Graduate School Fellowship (LGSF) at Emory University.

\section{Try on your own: Become a reverse-engineer}

By now we hope to have made a convincing case for our contention that different reverse-engineering methodologies are, in general, best-suited for answering different types of questions. We have reviewed the most prominent such questions, and illustrated how the ``goals'' fulfilled by specific algorithms are really manifestations of their underlying assumptions about what should count as an interaction. 

Since no one definition of biological interaction can be considered more ``correct'' than the others in all contexts (different algorithms merely capture different aspects of the same system), a diversity of goals and operational idiosyncracies might be viewed as a blessing rather than a curse. Yet choices should be made at the outset regarding what one wishes to learn by doing reverse-engineering, because these choices inform which algorithms are best suited for the job.

In this section, we simulate the conditions under which the need for such choices arises. Imagine that you have just been handed a set of high-throughput data, for a system whose interaction architectures have not yet been fully mapped. Follow the series of prompts in the box to embark on an exploratory challenge with a representative set of actual experimental data.

\vspace{0.1in}

\sloppy
Consider a set of multi-electrode recordings from the retina of a salamander (we thank M.~Berry for providing us with data from \cite{tkacik2014}; download link at~\url{https://figshare.com/articles/bint_fishmovie32_100_mat/5009840}). As explained in detail in the {\tt README.txt} file, the data consists of the responses from $p=160$ ganglion cells to the presentation of a naturalistic stimulus -- in this case, a short ($\sim 20$ sec) movie of a fish tank, repeated $n= 297$ times. The activity of each neuron is binarized as 0 (when the neuron is not firing an action potential) and 1 (when it is firing an action potential) within discrete time bins of length 20 ms. 

\begin{enumerate}

\item Of the methods discussed in this Chapter, which are clearly applicable to this particular set of data? Are there any which are not?

\item What kinds of predictions might a researcher want to make using this data?\vspace{0.05in}

\emph{Consider multiple levels of analysis, from single nodes in the neuronal network (Will removing a single node cause the network to collapse? Can we predict a future value for a given neuron, given the values of certain others?) to multiple nodes (Are there any functional groups that seem to be operating as a unit? Are there hub structures present?) to the entire system as an emergent whole (What can we say about the percentage of time the system is silent, versus when it is spiking? What other information would we need to say something about the ``typicality'' of the recorded networks, with respect to their structural and dynamical properties?).}

\item Crowdsourcing~\cite{surowiecki2004} -- the idea that conglomerate predictions, made by combining the wisdom of many independent thinkers, are more accurate than those of any individual -- is a popular strategy in DREAM competitions~\cite{marbach2012,saez2016} (for recent examples, see the closed Sage Bionetworks-DREAM Breast Cancer Prognosis (DREAM7, 2012), NIEHS-NCATS-UNC DREAM Toxicogenetics (DREAM8, 2013) and ICGC-TCGA DREAM Somatic Mutation Calling (DREAM 8.5-9, 2013-2014) Challenges). Yet we have seen that different reverse-engineering methods often yield disparate -- even antagonistic or contradictory -- predictions. For which combination of the following algorithms would you feel comfortable following the ``wisdom of crowds'' (say, averaging the results, or taking majority rules)?\vspace{0.05in}

\emph{Think about ARACNe, CLR, Bayesian networks (static and dynamics), MaxEnt approaches, and possibly other methods. Given the assumptions these methods make, would you take the union or intersection of the set of results produced by Bayesian methods and ARACNe? MaxEnt and CLR? Other combinations? When do you think crowdsourcing in general is a good strategy?}

\end{enumerate}

\newpage

\bibliography{Bibliography}

\end{document}